# SASE X-ray Free-Electron Laser driven by Resonant Multi-Pulse Ionization acceleration scheme

Federico Nguyen[1*], Paolo Tomassini[2], Leonida A. Gizzi[3], Luca Giannessi[4] and Anna Giribono[4]

[1]ENEA, Nuclear Department, I-00044, Frascati, Italy

[2]ELI-NP, R-077125, Magurele, Romania

[3]CNR-INO, I-56124, Pisa, Italy

[4]LNF-INFN, I-00044, Frascati, Italy

**Abstract.** Laser wakefield accelerators are on the way to provide GeV scale high-brightness electron beams for multidisciplinary applications. In the present work, electron bunches are studied in the framework of a Free-Electron Laser driven by an accelerator in the Resonant Multi-Pulse Ionisation framework. The transport to the undulator is provided by a matched beam focusing with marginal beam quality degradation. Finally, using the CRESCO/ENEAGRID High Performance Computing infrastructure, 3D time-dependent simulations of the produced radiation show that about $10^{10}$ photons with central wavelength of 0.15nm will be delivered with the electron bunches under consideration.

## 1 Introduction.

Nowadays, Free-Electron Laser (FEL) sources cover a wide range of applications, from materials science with effective technology transfer to industry to biomedical or pharmaceutical research advances. Efficient FEL operation in the X-ray domain requires the establishment of a high gain regime of energy transfer from the electron beam to X-ray photons, to finally yield a high-power laser pulse. This is achieved provided that the beam energy spread, and transverse emittances are sufficiently small to preserve the resonance condition set by the beam energy, the undulator period and deflection strength. The laser wakefield accelerator (LWFA) is a rapidly growing field of research and development in terms of potential flexibility and size reduction to drive an FEL. In the resonant multi-pulse ionization (ReMPI) injection [1], a tightly focused, short-wavelength pulse acts as the ionization pulse. Such a pulse extracts electrons from a dopant (e.g. N, Ar or Kr) and the remaining largest portion of the Ti:Sa pulse is time shaped as a sequence of sub-pulses and focused on the target with a long parabola. In this scheme, plasma density fluctuations result in the most severe source of beam quality fluctuations. Therefore, different electron bunches are simulated varying the plasma density by ±2.5% and by ±5% with respect to the reference condition. FEL performance is evaluated for each of these bunches, assuming to operate the source in self-amplified spontaneous emission (SASE) configuration, that allows the growth of spontaneous radiation power through exponential gain and saturation in the undulators section without any external laser seed. The stability of results upon variation of the starting noise seed is studied in detail for each bunch.

---

* Corresponding author. E-mail: federico.nguyen@enea.it

## 2 Transport of the electron beams accelerated in the ReMPI scheme

In this approach [1], the extracted electrons have small transverse momentum and size after the ionizing pulse passage, due to the pulse low amplitude and size. As a result, a 32pC charge bunch with a normalized emittance of $\epsilon_{nx}$=0.083mm×mrad along the ionizing pulse polarization is obtained at the capillary target end. Moreover, due to a partial overlapping between the ionizing pulse and the last pulse of the train, a quasi-round beam with normalized emittance $\epsilon_{ny}$=0.078mm×mrad is achieved. As the ionizing pulse is diffracted away at the capillary entrance, the driving train and the obtained trapped bunch enter the boosting section. Inside the energy boosting section, each sub-pulse (except for the first one) propagates into a plasma wave excited by the previous pulses of the train, so that the time evolution of each pulse is unique in the train and different from a standard evolution of a nonrelativistic pulse inside a capillary. At the end of the capillary density downramp, a ~ 4.5GeV electron beam with peak current of 3.5kA, projected energy spread of about 2% and slice energy spread at the current peak of $3\times10^{-4}$ is produced, with small differences depending upon variations of the plasma density.

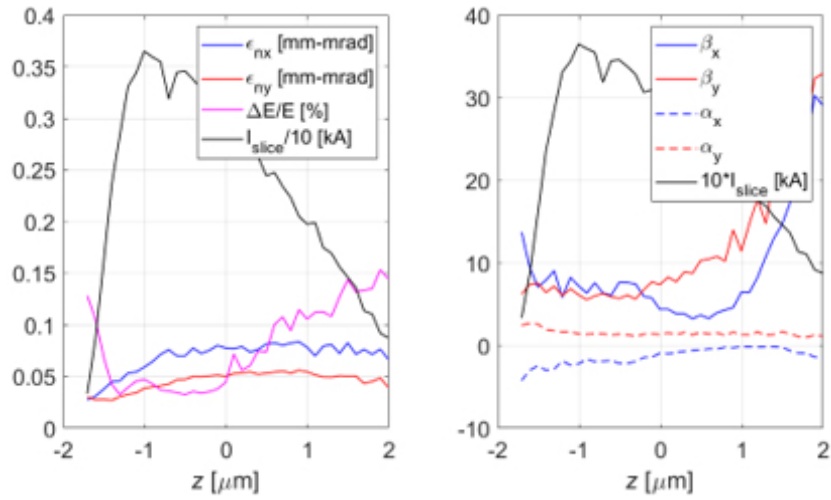

**Fig.1**: Slice analysis of the reference electron beam, at undulator entrance.

The beam transport and matching to the undulator line are enforced by a 10 m long transfer line consisting in the combination of state-of-the-art permanent and electromagnet quadrupoles. More in detail, the beamline is composed of three triplets of permanent magnet quadrupoles (PMQ) – the single triplet relying on 5-10-5 cm long quadrupoles with a nominal gradient of 450 T/m each spaced by 10 cm center-center – and four electromagnet quadrupoles (EMQ). The position and focusing gradients for PMQ quads have been settled aiming to capture the divergent beam taking advantage of the high magnetic field in a very short distance and trying to avoid spoiling the beam quality because of second order effects. The following EMQs, carry the beam at the undulator entrance and allow for Twiss parameters (α, β, and γ) tuning in a wide range acting only on their magnetic field. The transport beamline aims to carry the electron beam from the ReMPI exit stage and tackles the intrinsic transverse divergence and projected energy spread, potentially leading to emittance spoiling and chromaticity rise. In particular, the associated electron beam dynamics is explored by means of simulations with space-charge effects included, aiming to β values tunable in the range 8-12 m and absolute α functions around 1, with the beam converging in the vertical plane and diverging in the horizontal one. Figure 1 shows the slice analysis of the electron beam: the horizontal emittance along the beam always remains lower than 0.08mm×mrad values.

# 3 FEL beamline and performance

The undulator parameters chosen for the present study are the following: a planar undulator with undulator period $\lambda_u$=14mm and deflection parameter K=1.145, capable to operate at a gap of about 5mm and to provide a remanent magnetic field of 1.22T. With the electron bunches under consideration having an average energy of $E_{beam}$=4.5GeV, the expected resonant wavelength is $\lambda_R$≈0.15nm (namely about 8keV photon energy). The undulator beamline consists of 15 modules, each one with length $N_u$=140 periods, $L_u = N_u \times \lambda_u \approx$2m, so that the total active length is about 30 meters. With $E_{beam}$=4.5GeV a beam energy, the undulator has almost negligible focusing power, so that to keep the electron beam matched over the whole undulator beamline, the periodic magnetic cell has to include alternate gradient quadrupoles in between undulator modules. Accounting for fluctuations on the plasma density mainly affects the electron beam quality parameters at the undulator entrance and introduce a deviation from the nominal values of the Twiss functions of the beam entering the transport beamline. Four electron bunch distributions produced by varying the plasma density by ±2.5% and by ±5% with respect to the nominal condition, are carried through the same transport line and let interact and radiate through the same undulator beamline, in addition to the reference case.

Since the proposed FEL is designed to operate in SASE configuration, a relevant systematic effect is played by the initial noise power that fluctuates randomly. Therefore, the impact of the starting noise seed on the final performance is accounted for generating nine additional replicas of the FEL time-dependent simulation obtained by only varying the initial seed, for each bunch coming from the plasma exit, including the modified plasma density bunches. The FEL performance is studied with the GENESIS1.3 simulation code [2], by providing the related full 6D phase space. In particular, the code is used to perform 10 random initial seed simulations times 5 different plasma density samples, namely 50 different FEL simulations. Each simulation has the bunch split in 4500 independent longitudinal slices. The full propagation through the undulator beamline of each slice can be calculated independently in parallel, and it is performed on the CRESCO4 Cluster [3], after having compiled the code with the MPI IMPI-INTEL17 libraries. Results on relevant quantities are obtained by combining the output files resulting from the MPI data decomposition.

Each of the five panels of Figure 2 shows the growth of the pulse energy as a function of the undulator propagation coordinate, for a sample obtained with a different plasma density, specified in terms of fractional difference with respect to the reference value. Every highlighted curve is the average result out of the 4500 slices propagation, and out of 10 different noise seed simulations. On each panel, the highlighted curve is superimposed to the envelope of those 10 simulations and fit to the exponential distribution, for evaluating the gain length $L_g$ from the $e^{z/L_g}$ scaling behavior: upper left box shows fit results and errors. As a result, the gain length value $L_g = 1.8\ m$ is obtained, quite consistently among the five samples.

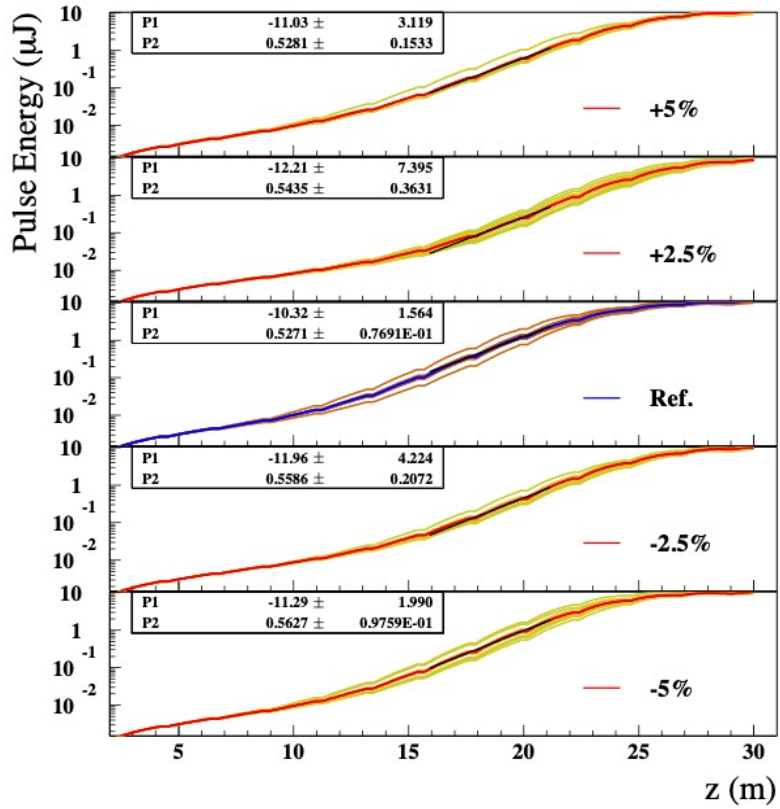

**Fig**.2: Pulse energy growth as a function of the undulator propagation coordinate for samples with different plasma density. The average growth is fit with an exponential function to estimate the gain length.

At undulator exit, the pulse energy is $E_p = 10.3\ \mu J$, that implies a flux of $4\times10^{23}$ photons/s and a brilliance of $3 \times 10^{31}\ photons\ s^{-1}\ (mm \times mrad)^{-2}\ (0.1\%\ bw)^{-1}$, with differences on the order of 9% among the different samples, with no significant trend as a function of plasma density.

In conclusion, by means of numerical simulations, a single-stage 4.5GeV LWFA scheme, along with a state-of-the-art beam transport line accommodating for the relatively large projected energy spread and large beam divergence, is proven to drive a SASE FEL source capable to deliver about $10^{10}$ photons/shot at 0.15nm resonant wavelength, at the exit of an undulator beamline 30m long.